\documentclass[12pt]{article}
\usepackage{latexsym}
\usepackage{amsmath}
\usepackage{amsfonts}
\usepackage{amssymb}
\usepackage{graphicx}

\usepackage{color}
\usepackage{tikz}
\usepgflibrary{arrows}

\newcommand{\mR}{{\mathbb R}}

\newcommand{\mA}{{\mathcal A}}
\newcommand{\mD}{{\mathcal D}}

\title{Localizability, gauge symmetry and Newton-Wigner operator for massless particles}
\author{P.Kosi\'nski\thanks{e-mail: pkosinsk@uni.lodz.pl},\;
P.Ma\'slanka  \\
\small Department of  Computer Science, 
 Faculty of Physics and Applied Informatics\\
\small University of \L\'od\'z,\\
\small Pomorska 149/153, 90-236 {\L}\'od\'z, Poland
}
\date{}
\begin{document}
\maketitle 
\begin{abstract}
The notion of position operator for massless spinning particles is discussed in some detail. The noncommutativity of coordinates is related to the gauge symmetry following from the freedom in definition of standard state in Wigner's construction of induced representations of Poincare group. The counterpart of Newton-Wigner operator is discussed. It is explained why the Newton-Wigner construction works only for helicity $ \mid\lambda\mid\leq1/2$.
\end{abstract}
\section{Introduction}
Particle localizability is a well defined notion in nonrelativistic quantum mechanics.  However, it lacks the precise meaning in relativistic (quantum) regime. The crucial property of relativistic quantum theory is the absence of nontrivial  interactions admitting particle number conservation. This has to be carefully taken into account when particle localization procedures are analyzed\cite{b1}. The main restriction comes from the fact we are dealing with a single particle while the one-particle relativistic quantum theory is only an approximation valid, for example, for weak and slowly varying external fields. However, once the validity of such approximation is assumed one can construct a number of candidates for particle position operator.

\par
The most reasonable proposal has been given by Newton and Wigner in their seminal work \cite{b2}, anticipated to some extent by Price \cite{b3}. The localizability problem has been further discussed in a number of papers \cite{b4}$\div$\cite{b10}, to mention only a few.
\par
One can argue that any reasonable position operator should have the following properties:
\begin{itemize}
\item[(i)] one would like to retain the interpetation of momentum operator $\hat{\vec{P}}$ as generator of translations in space; as a result the following commutation rule is assumed \\
$$ [\hat{X_i}, \hat{P^k}]=i\delta^k_i   ;$$ \\
\item[(ii)]the coordinate operator $ \hat{X_i} $ should transform as a threevector under rotations; \\
\item[(iii)] the existence of coordinate representation is assumed which implies the commutation rules
$$ [\hat{X_i}, \hat{X_k}]=0$$
\end{itemize}
It was realized some time ago that one can hardly expect all the above conditions to be fulfilled in 
the case of spinning massless particles. In fact, already Newton and Wigner noticed that their construction fails to work in the limit of vanishing mass, except for helicities $ \lambda=0 $ and $ \mid\lambda\mid=1/2 $ ( in the latter case both helicities must be included). A simple explanation of this fact is essentially contained in a very nice paper by Skagerstam \cite{b11}. Basically, the reason that Newton-Wigner algorithm does not work for $ \mid\lambda\mid>1/2 $ is that, given the position operator obeying (i) $ \div $ (iii), one can construct $ SU(2) $ spin variables which is, in turn, only possible if all helicities $-\mid \lambda \mid, -\mid \lambda \mid+1,...,\mid \lambda \mid-1, \mid \lambda \mid $ are included. This Mukunda-Skagerstam argument is explained in some detail in Sec. II. In Sec. III we sketch the construction of representations of Poincare group for massless particles from the point of view of gauge symmetry following from the freedom in the choice of standard state used in Wigner's construction of induced representations. Such an approach  provides a convenient framework for discussing the problem of noncommutativity of coordinates. Sec. IV is devoted to the discussion of Newton-Wigner coordinate for parity conserving helicity $1/2$ massless theory. We show that in this special case the coordinate operator can be identified with covariant ( with respect to the gauge symmetry mentioned above ) derivative which describes trivial connection. This is a consequence of dealing with reducible representation of Poincare group encompassing both helicities. The nontrivial connections corresponding to both helicities are obtained from the trivial one by abelian projection. In Sec. V photons as an example of  $ \mid \lambda \mid=1 $ are considered. Again the covariant derivatives for both helicities are obtained by abelian projection from $ SU(2) $ connection  which, in contrast to $ \mid \lambda \mid=1/2 $ case, is also nontrivial; its nontriviality is related to the transverality  condition, i.e. lacking of $ \lambda=0 $ helicity component ( in accordance with Mukunda-Skagerstam argument ). The nontriviality of the connection results in breaking (iii) for photon coordinate operator. Sec. VI is devoted to some conclusions. In Appendix we discuss the source of nonminimal (from the point of view of gauge symmetry) terms in Poincare generators. We sketch also the construction of position operators which obey (i) and (iii) but transform nonlinearly under rotations.

\section{Localizability of massless particles}

Consider a free particle described by an irreducible representation of Poincare group. The group generators $P_\mu$ ( fourmomentum ), $\vec{J}$ (angular momentum ) and $\vec{K}$ (boosts ) are represented by observables acting in the space of states and obeying the algebra 

\begin{equation}
\label{e1}
\begin{split}
&[J_i,J_k]=i\varepsilon_{ikl}J_l,\quad \quad\;\;[J_i,P_k]=i\varepsilon_{ikl}P_l,\\
&[J_i,K_k]=i\varepsilon_{ikl}K_l , \quad \quad[K_i,P_k]=i\delta_{ik}P_0, \\
&[K_i,K_k]=-i\varepsilon_{ikl}J_l ,  \quad \;[K_i,P_0]=iP_i
\end{split}
\end{equation}
with the remaining commutators vanishing.\\
Assume one can construct the observable $\vec{X}$ representing the particle position. The basic conditions $(i)\div(iii)$ from Introduction imply the following commutation rules: 

\begin{equation}
\label{e2}
\begin{split}
&[X_i,P^k]=i\delta_i^k\\
&[J_i,X_k]=i\varepsilon_{ikl}X_l\\
&[X_i,X_k]=0
\end{split}
\end{equation}
Let us define the orbital angular momentum operator

\begin{equation}
\label{e3}
L_i\equiv\varepsilon_{ijk}X_jP^k
\end{equation}
together with the spin operator
\begin{equation}
\label{e4}
S_i \equiv J_i-L_i
\end{equation}
Eqs. $(1)\div(4)$ imply

\begin{equation}
\label{e5}
\begin{split}
&[S_i,S_k]=i\varepsilon_{ikl}S_l\\
&[S_i,P_k]=0
\end{split}
\end{equation}

Therefore, our space of states carries some representation of $sU(2)$ algebra spanned by the spin operators $S_i$.
\par
Consider now the case of massless particles. The relevant representations of Poincare group are uniquely determined by the choice of (half)integer helicity $\lambda$ defined by 

\begin{equation}
\label{e6}
\lambda \equiv \frac{\vec{P}}{\mid \vec{P} \mid}\cdot\vec{J}=\vec{S}\cdot\frac{\vec{P}}{\mid \vec{P} \mid}
\end{equation}

Due to the second eq.(\ref{e5}) one can diagonalize $\vec{S}\cdot\frac{\vec{P}}{\mid \vec{P} \mid}$ within any subspace carrying irreducible representation of spin algebra. Therefore, the spectrum of $\vec{S}\cdot\vec{P}$ contains, together with $\lambda$, all values $-|\lambda|, -|\lambda|+1,...,|\lambda|-1,|\lambda|$.\\
This is, however, in contradiction with the operator relation (\ref{e6}). This nice argument has been given by Skagerstam \cite{b11} who ascribed it to N. Mukunda.
\par
It is now quite easy to understand why Newton-Wigner scheme works in the massless case only for $|\lambda|=0,1/2$. The spinless case is trivial. For $|\lambda|=1/2$, the values $\lambda=\pm1/2$ cover the whole spectrum of $\vec{S}\cdot\frac{\vec{P}}{\mid \vec{P} \mid}$. Therefore, there is no contradiction provided we consider the reducible representation including both helicities which is the case for Dirac equation considered in \cite{b2}. For $|\lambda|>1/2$ one would be forced to consider the irreducible representations of Poincare group describing particles of different absolute values of helicities.

\section{Massless representations from gauge principle}
\label{s3}
The construction of irreducible representations of Poincare group was described in the seminal paper of Wigner \cite{b12} and repeated in numerous articles and books. It is technically slightly involved; in the case of massless particles additional difficulty comes from the fact that the relevant orbit in momentum space is topologically nontrivial.
\par
One can look at the problem of massless representations from slightly different point of view. We start with the simplest representation carried by the complex valued functions $f(\vec{p})$ equipped with a scalar product
\begin{equation}
\label{e7}
(f,g)\equiv\int\frac{d^3\vec{p}}{p^0}\overline{f(\vec{p})}g(\vec{p}), \quad p^0=|\vec{p}|
\end{equation}
The Poincare group acts by $f(\vec{p})\rightarrow f(\Lambda^{-1}\vec{p})e^{ipa}$ which yields irreducible representation corresponding to vanishing mass and helicity. The group generators read

\begin{equation}
\label{e8}
\begin{split}
&P_\mu=p_\mu\\
&J_i=i\varepsilon_{ijk}\frac{\partial}{\partial p^j}p^k \\
&K_i=\frac{1}{2}\left(p^0(i\frac{\partial}{\partial p^i}- \frac{i}{2}\frac{p^i}{(p^0)^2}) + 
(i\frac{\partial}{\partial p^i}- \frac{i}{2}\frac{p^i}{(p^0)^2})p^0\right)=ip^0\frac{\partial}{\partial p^i}
\end{split}
\end{equation}

Consider now the case of nonvanishing helicity, $\lambda\neq0$. One selects some standard fourvector 
$\underline{k}$ (typically, $\underline{k}=(k,0,0,k)$) and standard boost $L(p)\in SO(3,1)$ obeying $L^\mu_{\;\;\nu}(p)k^\nu=p^\mu$.The basic vectors are defined by

\begin{equation}
\label{e9}
|p \rangle = U(L(p))|\underline k \rangle 
\end{equation}
where $|\underline k \rangle $  spans one dimensional holomorphic representation of stability subgroup of $\underline k$ ( which is isomorphic to $E(2)$). The choice of the standard boost is not unique. One can make replacement $L(p)\rightarrow L(p)L_0(p)$, where $L_0(p)$ is any element of stability subgroup of $\underline k , L_0(p)\underline k =\underline k $. It is crucial to note that $L_0$ may depend on final momentum $\vec{p}$. The corresponding modified basic vector reads

\begin{equation}
\label{e10}
\begin{split}
|p \rangle' = &U(L(p)L_0(p))|\underline k \rangle=U(L(p))U(L_0(p))|\underline k \rangle = \\
&e^{i \lambda\varphi(p)}U(L(p))|\underline k \rangle=e^{i \lambda\varphi(p)}|\underline p \rangle
\end{split}
\end{equation}

which induces the gauge transformations for amplitudes $f(\vec{p}) \equiv \left\langle \vec{p} |f\right\rangle $: 
\begin{equation}
\label{e11}
f'(\vec {p}) = e^{-i \lambda\varphi(\vec {p})}f(\vec {p})
\end{equation}
This gauge transformation corresponds merely to the change of conventions so it yields unitarily equivalent representation; in particular, the structure of Lie algebra is invariant under the gauge transformations(\ref{e11}). This implies that one has to replace the derivatives $i\frac{\partial}{\partial p^k} \equiv i\partial_k$ entering the generators (\ref{e8}) by their covariant counterparts \cite{b11}
\begin{equation}
\label{e12}
i{\mD}_k\equiv i \partial_k-{\mA}_k ,
\end{equation}

where the gauge field ${\mA}_k$ transforms as follows

\begin{equation}
\label{e13}
{\mA}'_k(p) = {\mA}_k(p) +\lambda \partial_k \varphi(p)
\end{equation}
It remains to determine the gauge class of ${\mA}_k(p)$. This problem can be solved by taking a closer look at Wigner construction (for example, by comparing the action of stability subgroups corresponding to opposite standard (three)momenta). As a result one concludes that the massless particles of helicity $\lambda$ are describes by line bundle over the sphere $S^2 (|\vec{p}|=$ const)
characterized by the integer $2\lambda$. Noting additionally that the Lie algebra commutation rules should be explicitly covariant under rotations one finds that ${\mA}_k$ corresponds to the field strength of monopole with "magnetic" charge $\lambda$  \cite{b11},\cite{b22}, \cite{b13},\cite{b23} . However, this is not the end of the story.  The "minimal" coupling $i\partial_k \rightarrow i{\mD}_k$ yields generators still obeying $\vec{P}\cdot\vec{J}=0$. Therefore, the definition of $\vec{J}$ must be supplied by nonminimal term producing the correct relation (\ref{e6}). Finally one obtains 

\begin{equation}
\label{e14}
\begin{split}
&J_i=i\varepsilon_{ijk}D_jp^k + \lambda \frac{p^i}{|\vec{p}|}\\
&K_i=\frac{1}{2}\left(p^0(iD_i- \frac{i}{2}\frac{p^i}{(p^0)^2}) + 
(iD_i- \frac{i}{2}\frac{p^i}{(p^0)^2})p^0\right)=ip^0 D_i
\end{split}
\end{equation}
One easily checks that the choice (\ref{e14})  yields correct commutation rules. 
\par
Coming back to the definition of coordinate operator $X_i$  note that the first equation  (\ref{e2}) implies that $X_i$ must contain $i\partial_i$. Therefore, demanding that $X_i$ does not depend on the choice of conventions ( the gauge choice ) we conclude that $i\partial_i$ enters coordinates through covariant derivative $iD_i$ \cite{b11}.The natural choice is simply $X_i= i{\mD}_k$ Then the coordinates necessarily do not commute. Taking into account that the potential $\vec{\mA}(\vec{p})$ describes a monopole of charge $\lambda$ one immediately finds

\begin{equation}
\label{e15}
[X_i, X_j] = -i\lambda \varepsilon_{ijk} \frac{p^k}{(p^0)^3}
\end{equation}
Therefore, (iii) is not fulfilled. Let us note that the non-commutativity of the position operator can be derived from the classical description of spinning particles \cite{b24}, \cite{b25}.
\par
 On the other hand, one finds that $X_i$'s transform as the components of threevector,  
\begin{equation}
\label{e16}
[J_i, X_j] = i\varepsilon_{ijk}X_k
\end{equation}
so (ii) holds. This is slightly surprising because one cannot find explicitly rotationally invariant potentials describing the field of monopole. It is interesting to consider this problem in some detail. The description of symmetries in gauge theories demands some care \cite{b14}. In fact, a given transformation is a symmetry if its "unwanted" effect can be cancelled by a suitable compensating gauge transformation. In our case the change of vector potential under rotations is cancelled by $U(1)$ gauge transformation (see Appendix for some details). As a result the initial generators must be supplied  by additional pieces which generate the relevant compensating gauge transformations ( this is the origin of well known "spin from isospin" phenomenon \cite{b15}). In the case under consideration 
the second term on the right hand side of the first eq. (\ref{e14}) may be viewed  as a generator of compensating gauge transformation.
\par
Obviously, the choice $X_i=i{\mD}_i$ is not unique. If one insists on commuting coordinates, $X_i$ must be modified; then (ii) is no longer fulfilled. An example of such modified coordinate is considered in Ref. \cite{b16}. The coordinate observables are there obtained by quantizing Darboux coordinates on coadjoint orbit of Poincare group; obviously, they do commute. The price to pay is that they transform nonlinearly under rotations which linearize on $SO(2)$ subgroup only.  Some details are given in Appendix.

 \section{Newton-Wigner coordinate }

As we have seen in Sec.II the only case one can avoid the general conclusions of Mukunda and Skagerstam ( apart from the trivial case of scalar particles ) is $|\lambda|=1/2$ provided both helicities are taken into account. The dynamics (or rather kinematics) of such particles is described
by free massless Dirac equation:
 
\begin{equation}
\label{e17}
i\frac{\partial \Psi}{\partial t} = \vec{\alpha}\vec{p}  \Psi       
\end{equation}

We work with the following representation of Dirac  matrices:

\begin{equation}
\label{e18}
\vec{\alpha}=
\left(
\begin{matrix}
\vec{\sigma} &0\\
0& - \vec{\sigma} 
\end{matrix} 
\right),
 \quad
\beta=\left(
\begin{matrix}
0 &I\\
I&0 
\end{matrix} \right)   
\end{equation}
The general positive energy solution to eq. (\ref{e17}) reads

\begin{equation}
\label{e19}
\Psi(x)=\frac{1}{(2\pi)^{3/2}}\int d^3\vec{p} \left[ 
\left( \begin{matrix}
u_+(p)\\
0
\end{matrix} 
\right)  
a_+(p)+
\left( \begin{matrix}
0\\
u_-(p)
\end{matrix} 
\right)
a_-(p
\right]
e^{i\vec{p}\vec{x} - ip^0t}   
\end{equation}
where $u_\pm(p)$ are spinors obeying

\begin{equation}
\label{e20}
\begin{split}
&p^k\sigma_k u_\pm(p)=\pm p^0u_\pm(p) \\
&u_\pm(p)^+u_\pm(p)=1, \quad u_\pm(p)^+u_\mp(p)=0
\end{split}
\end{equation}
One possible choice is

\begin{equation}
\label{e21}
\begin{split}
u_+(p)=\frac{1}{\sqrt{2p^0(p^0-p^3)}} 
\left( \begin{matrix}
p^1-ip^2)\\
p^0-p^3
\end{matrix} 
\right)  \\  
u_-(p)=\frac{1}{\sqrt{2p^0(p^0+p^3)}} 
\left( \begin{matrix}
-p^1+ip^2)\\
p^0+p^3
\end{matrix} 
\right)
\end{split}    
\end{equation}
The scalar product reads

\begin{equation}
\label{e22}
\int d^3 \vec{x} \Psi^+(\vec{x},t) \widetilde {\Psi}(\vec{x},t) = \int d^3 \vec{p} \left[ \overline{a}_+(p) \widetilde{a}_+(p) + \overline{a}_-(p) \widetilde{a}_-(p) \right]  
\end{equation}

It differs from (\ref{e7}) by a simple redefinition of amplitudes and makes $i\partial_k$ (and $iD_k $) 
hermitean at the expense of slightly complicating Poincare transformations rules.\\
The spaces of the amplitudes $ a_\pm(p) $ carry helicity $ \lambda=\pm\frac{1}{2} $ representations. The $ m=0 $ limit of Newton-Wigner operator \cite{b2}
, \cite{b3} reads 

\begin{equation}
\label{e23}
X_k=x_k + \frac{i}{2} \left( \frac{\beta \alpha^k}{p^0}+\frac{(\vec{p}\vec{\alpha})\alpha^k-p^k}{(p^0)^2}- \frac{\beta(\vec{p}\vec{\alpha})p^k}{(p^0)^3}\right) 
\end{equation}

with $ x_k $  and $ p^k $ having standard meaning.\\
Acting on the wave function (\ref{e19}) one finds that $ X_k $ preserves the positivity of energy, i.e. the result is of the same form (\ref{e19}) with new amplitudes $ a_\pm (p)$ given by

\begin{equation}
\label{e24}
\left(\begin{matrix}
a'_{+k}(p)\\
a'_{-k}(p)
\end{matrix}\right)=
\left[ i\partial_k + i 
\left( \begin{matrix}
u^+_+ \partial_k u_+ & u^+_+ \partial_k u_-\\
u^+_- \partial_k u_+ & u^+_- \partial_k u_-
\end{matrix}\right) 
\right]
 \left(\begin{matrix}
a_{+k}(p)\\
a_{-k}(p)
\end{matrix}\right) 
\end{equation}
We see that the Newton-Wigner coordinate operator takes the form

\begin{equation}
\label{e25}
X_k=i{\mD}_k \equiv i \partial_k - {\mA}_k
\end{equation}

\begin{equation}
\label{e26}
 {\mA}_k \equiv (-i) \left( \begin{matrix}
u^+_+ \partial_k u_+ & u^+_+ \partial_k u_-\\
u^+_- \partial_k u_+ & u^+_- \partial_k u_-
\end{matrix}\right) 
\end{equation}

and can be viewed as $ U(2) $ covariant derivative. \\

The non diagonal elements are nonvanishing so, as expected, both helicities are needed to define Newtom-Wigner coordinate. The diagonal elements are the monopole vector potentials describing monopoles of charges $ \pm 1/2 $ . They represent the potentials  corresponding to the Dirac strings extending along positive and negative third semiaxis, respectively. One can convert $ {\mA}_k $ into $ sU(2) $ element by extracting the trace which is a pure gauge except singularity along third axis. Then the diagonal elements of modified $ {\mA}_k $ describe monopoles in the gauge involving two strings along both positive and negative semiaxis, each carrying half of the total flux. However, for simplicity we will be dealing with $ U(2) $ connection given by (\ref{e26}).\\
The Newton-Wigner coordinates commute which implies that $ {\mA}_k $ should be a pure gauge. Indeed, it is immediately seen that 
\begin{equation}
\label{e27}
 {\mA}_k =-i\mathcal{U}^+ \partial_k \mathcal{U}
\end{equation}
with
\begin{equation}
\label{e28}
 \mathcal{U}=\left( \begin{matrix}
 (u_+)_1 & (u_-)_1 \\
 (u_+)_2 & (u_+)_2
 \end{matrix}\right)
\end{equation}

Let us note that the nontrivial monopole connections describing irreducible representations of definite parity result from abelian connection (\ref{e27}) \cite{b17}; this is a counterpart of 'tHooft mechanism \cite{b18}.

\;\;
The gauge potential transforms according to the rule $ {\mA}'_k=V{\mA}_k V^+ -iV\partial_kV^+ $.
Choosing $ V=\mathcal{U}^+ $ we find $ {\mA}'_k=0 $. Therefore, in the new basis the coordinate operator
$ \hat{X}_k $ equals simply $ i\partial_k - \frac{i}{2}\frac{p^k}{(p^0)^2} $ (which is the partial derivative hermitean with respect to the scalar product (\ref{e7})). New generators of Lorentz group read
\begin{equation}
\label{e29}
\begin{split}
 &J_i= -i \varepsilon_{ijk} p^j \partial_k + \frac{1}{2} \sigma_i \\
 &K_i= i p^0 \partial_i + \frac{1}{2p^0} \varepsilon_{ijk}p^j \sigma_k
\end{split}
\end{equation}

 \section{The case $|\lambda|=1$ }

For completeness we discuss now the case $|\lambda|=1$ (photons). According to the arguments of Mukunda and Skagerstam, presented in sec. (II), for $ |\lambda|>1/2 $ it is not possible to find a coordinate operator obeying $(i) \div (iii) $, even if both helicities $ \pm |\lambda| $ are included. As an example, consider the case of photons, discussed already by Skagerstam \cite{b11}.\\
In the standard formalism \cite{b19} the photon wave function can be written as 
\begin{equation}
\label{e30}
f_{\mu\nu}= \int \frac{d^3\vec{p}}{(2\pi)^32p^0}\left( e_{\mu\nu}(\vec{p})\varphi(\vec{p},1)+\overline{e_{\mu\nu}(\vec{p})}\varphi(\vec{p},-1)\right) e^{-ipx}
\end{equation}
with $ e_{\mu\nu}(\vec{p}) $ being the polarization tensor. Momentum amplitudes $ \varphi(\vec{p},\pm 1) $ span the massless representations of Poincare group corresponding to helicities $ |\lambda|=\pm 1 $. The scalar product in the space of states carrying both helicities (reducible representation ) reads
\begin{equation}
\label{e31}
 (\varphi,\psi)=\int \frac{d^3\vec{p}}{(2\pi)^32p^0}\left(\overline{\varphi(\vec{p},1)}\psi(\vec{p},1) + \overline{\varphi(\vec{p},-1)}\psi(\vec{p},-1)\right)
\end{equation}
Transformation rules for amplitudes $ \varphi(\vec{p},\lambda) $ read
\begin{equation}
\label{e32}
 \varphi(\vec{p},\lambda) = e^{i\lambda\Theta(p,\Lambda)}e^{i\vec{p}\vec{a}} \varphi(\overrightarrow{\Lambda^{-1}p},\lambda)
\end{equation}
The polarization tensor $  e_{\mu\nu}(\vec{p}) $ ( intertwinning operator ) can be expressed in terms of polarization vector $ \vec{e}(\vec {p}) $ \cite{b19}
\begin{equation}
\label{e33}
\begin{split}
 &e_{0i} = ip_0e_i(\vec{p})\\
 &e_{ij} = -p_0 \varepsilon_{ijk} e_k(\vec{p})
 \end{split}
\end{equation}

 obeying
 
 \begin{equation}
 \label{e34}
\begin{array}{ll}
\vec{p}\cdot\vec{e}(\vec{p})=0,\quad \quad & \vec{p}\times \vec{e}(\vec{p})= -i p_0 \vec{e}(\vec{p}) \\
 \vec{e}(\vec{p}) \cdot \vec{e}(\vec{p})=0,\quad \quad &  \overline{\vec{e}(\vec{p})} \cdot \vec{e}(\vec{p})=1 \\
\overline{ \vec{e}(\vec{p})}\times \vec{e}(\vec{p}) = i\frac{\vec{p}}{p_0} &
 \end{array}
\end{equation}

Let us define   
   
 \begin{equation}
 \label{e35}
\vec{\varphi}(\vec{p})= \vec{e}(\vec{p})\varphi(\vec{p},1) + \overline{\vec{e}(\vec{p})}\varphi(\vec{p},-1)
\end{equation} 
 
 Then  
   
 \begin{equation}
 \label{e36}
\int \frac{d^3\vec{p}}{(2\pi)^32p^0}\overline{\vec{\varphi}(\vec{p})}\cdot\vec{\psi}(\vec{p}) = \int \frac{d^3\vec{p}}{(2\pi)^32p^0}\left(\overline{\varphi(\vec{p},1)}\psi(\vec{p},1) + \overline{\varphi(\vec{p},-1)}\psi(\vec{p},-1)\right)
\end{equation}    

and

\begin{equation}
 \label{e37}
\vec{p}\cdot \vec{\varphi}(\vec{p}) = 0
\end{equation} 
   
This is the form of representation discussed in Ref. \cite{b11}. The space of states is the transverse subspace of the Hilbert space of vector valued square integrable functions. The generators of the Poincare group consist of fourmomenta $ p^\mu $ together with the Lorentz group generators

\begin{equation}
 \label{e38}
\begin{split}
J_k& = \varepsilon_{klm}\left( i\partial_l - \frac{i}{2}\frac{p^l}{(p^0)^2}-{\mA}^a_l(\vec{p})\mathcal{S}_a\right)p^m \\
K_K &= \frac{1}{2}\left(p^0(i\partial_k-\frac{i}{2}\frac{p^k}{(p^0)^2}-{\mA}^a_k(\vec{p})\mathcal{S}_a) +
(i\partial_k-\frac{i}{2}\frac{p^k}{(p^0)^2}-{\mA}^a_k(\vec{p})\mathcal{S}_a)p^0\right)\\
& =  ip^0\partial_k -ip^0{\mA}^a_k(\vec{p})\mathcal{S}_a
\end{split}
\end{equation} 

where

\begin{equation}
 \label{e39}
(\mathcal{S}_a)_{kl}=-i\varepsilon_{akl}
\end{equation} 

is the spin generator and

\begin{equation}
 \label{e40}
{\mA}^a_k(\vec{p})=\varepsilon_{alk}\frac{p^l}{(p^0)^2}
\end{equation} 

is the Wu-Yang monopole \cite{b20}. Note that $ {\mA}^a_k(\vec{p}) $ is regular on $ {\mR}^3 \smallsetminus \{ 0 \} $; this is related to the fact that principal $ SU(2) $ bundles over $ S^2 $ are trivial.\\
We see that again the generators are expressed in terms of covariant derivative ( this time corresponding to $ SU(2) $ connection):

\begin{equation}
 \label{e41}
({\mD}_k)_{lm}=i\delta_{lm}\partial_k - {\mA}^a_k(\mathcal{S}_a)_{lm}
\end{equation} 

To make the contact with $ U(1) $ gauge symmetry described previously let us note that, by virtue of eqs. (\ref{e34}), the polarization vector $ \vec{e}(\vec{p}) $ can be written in the form \cite{b19}

\begin{equation}
 \label{e42}
\vec{e}(\vec{p})= \frac{1}{\sqrt{2}}\left(\vec{l}_1(\vec{p}) + i\vec{l}_2(\vec{p})\right)
\end{equation} 

where $\vec{l}_1  $ and $ \vec{l}_2 $ obey

\begin{equation}
 \label{e43}
\vec{l}_1(\vec{p}) \times \vec{l}_2(\vec{p})=\frac{\vec{p}}{p^0}, \quad \quad \vec{l}_2(\vec{p}) \times \frac{\vec{p}}{p^0}= \vec{l}_1(\vec{p}),\quad \quad \frac{\vec{p}}{p^0} \times \vec{l}_1(\vec{p}) = \vec{l}_2(\vec{p})
\end{equation} 

The set $  \{\vec{l}_1(\vec{p}), \vec{l}_2(\vec{p}) \}$ is defined up to an arbitrary rotation around $ \vec{p} $ by an  angle $ \omega (\vec{p}) $. This results in the following transformation of 
$ \vec{e}(\vec{p}) $:

\begin{equation}
 \label{e44}
\vec{e}(\vec{p}) \rightarrow  e^{i\lambda\omega (\vec{p})}\vec{e}(\vec{p})
\end{equation} 

which corresponds to a gauge transformation

\begin{equation}
 \label{e45}
\varphi(\vec{p},\lambda) \rightarrow  e^{-i\omega (\vec{p})}\varphi(\vec{p},\lambda)
\end{equation}

Again, as in the fermionic case, we are dealing with $ U(1) $ gauge symmetry corresponding to $ \lambda=\pm1 $ charges. In the original subspaces spanned by $ \varphi(\vec{p},\pm1) $ the Poincare generators are the functions of momenta and $ U(1) $ covariant derivatives

\begin{equation}
 \label{e46}
{\mD}_k^{\lambda} \equiv i\partial_k - {\mA}^{\lambda}_k
\end{equation}

where $ {\mA}^{\lambda}_k $ are the ( singular on $ {\mR}^3 \smallsetminus \{ 0 \} $ ) potentials of $ U(1) $ monopoles,

\begin{equation}
 \label{e47}
\partial_l {\mA}^{\lambda}_k-\partial_k {\mA}^{\lambda}_l = \lambda \varepsilon _{klm} \frac{p^m}{(p^0)^3}
\end{equation}

In order to find the relation between $ {\mD}_k $ and $ {\mD}_k^{\lambda} $ we write the consistency condition

\begin{equation}
 \label{e48}
i({\mD}_k)_{mn}\varphi_n(\vec{p}) = e_m(i\partial_k - {\mA}^{(+)}_k)\varphi(\vec{p},1) + 
\overline{ e_m}(i\partial_k - {\mA}^{(-)}_k)\varphi(\vec{p},-1)
\end{equation}

which yields

\begin{equation}
 \label{e49}
 \begin{split}
{\mA}^{(+)}_k=-i\overline{ e_m}\partial_k e_m = \vec{l}_1\cdot\partial_k\vec{l}_2 \\
{\mA}^{(-)}_k=-ie_m \partial_k \overline{ e_m} = -\vec{l}_1\cdot\partial_k\vec{l}_2 
\end{split}
\end{equation}

It is straightforward to check that eqs. (\ref{e49}) imply (\ref{e47}). We conclude that passing from the representation in terms of amplitudes $ \varphi(\vec{p},\pm1) $ to that based on $ \vec{\varphi}(\vec{p}) $
implies embedding the singular $ U(1) $ connection into regular $ SU(2) $ one.\\
Now, one can attempt to define the coordinate operator by

\begin{equation}
 \label{e50}
 \hat{X}_k = i {\mD}_k
\end{equation}

Then (i) and (ii) are fulfilled. However, the components of $  \hat{X}_k $ do not commute due to the nontriviality  monopole potential (\ref{e40}); instead one finds \cite{b11}

\begin{equation}
 \label{e51}
 [\hat{X}_k, \hat{X}_l] = -i \varepsilon_{klm} \frac{p^m}{(p^0)^4} p^a \mathcal{S}_a
\end{equation}

so (iii) is broken. This is due to lack of third helicity value, $ \lambda=0 $. Indeed, by virtue of

\begin{equation}
 \label{e52}
( \mathcal{S}_a\frac{p^a}{p^0})^2_{lm}=\delta_{lm} - \frac{p^lp^m}{(p^0)^2}
\end{equation}

one finds that the transversality condition (\ref{e37}) excludes $ \lambda=0 $. It is, however, the
transversality condition which makes both $ U(1) $ and $ SU(2) $ connections nontrivial. This is in contrast with $ \mid \lambda \mid = \frac{1}{2} $ case where the abelian projection yields nontrivial
 $ U(1) $ connection from trivial $ SU(2) $ one.
 
\section{Conclusions }

We have shown that, for the massless particles, $ \mid \lambda \mid = \frac{1}{2} $ is the only case ( apart from the trivial $ \mid \lambda \mid = 0 $ one )where the Newton-Wigner procedure works yielding the coordinate operator with commuting components. This is because taking massless Dirac equation one obtains reducible representation of Poincare group  comprising both helicities, $ \lambda=\pm \frac{1}{2} $. 
Then the space of states carries twodimensional representation of $ SU(2) $ spin variables which, by the Mukunda-Skagerstam theorem, is a necessary condition for the existence of coordinate operator obeying natural conditions (i) $ \div $ (iii) listed in the Introduction. For higher helicities some states are lacking and the relevant $ SU(2) $ representation cannot be accommodated; the acceptable
( from the point of view (i) $ \div $ (iii) ) coordinate operator does not exist. From the technical point of view the lack of some helicities imposes the constraints on momentum amplitudes which makes impossible to embed the initial nontrivial $ U(1) $ connections into trivial nonabelian one. The appearance of covariant derivatives (and, consequently, the related connections ) is indispensable 
for the representation of Poincare group to be independent of the convention used to define the standard state in the framework of induced representations.
\par
One can, of course, weaken the conditions   (i) $ \div $ (iii) resigning, for example, from explicit 
rotation covariance ( the condition (ii) ). Then the coordinate operator can be constructed for any $ \lambda $, even within an irreducible representation. To this end one can start from Darboux coordinates on coadjoint orbit of Poincare group \cite{b16}. The resulting Hamiltonian system, upon quantization, yields the relevant irreducible representation of Poincare group. In particular, one obtains commuting coordinates. However, the price one has to pay is that they transform under rotation according to nonlinear realization linearizing  only on $ SO(2) $ subgroup of rotation group. An attempt to linearize the transformation rule leads to noncommuting coordinates. Some details are given in Appendix.
\par
Let us point out that the problem of chiral particles as classical massless spinning particles has been studied in the symplectic framework in a very interesting paper by Duval and Horvathy\cite{b21}
\section{Appendix }

\subsection{Nonminimal terms and rotational invariance}

As we have noted in Sec. III the correct form of rotation generators is obtained only after including an  additional term which is not generated by minimal coupling principle. Its appearance is related to the fact that there exists no rotationally invariant vector potential describing the magnetic monopole field. Here we discuss this issue in some details.
\par
A vector field transforms under rotations as follows

\begin{equation}
 \label{e53}
{\mA}'_i(p')=O_{ij}{\mA}_j(p), \quad \quad \quad p'^k=O_{kl}p^l
\end{equation}

where $O$ is orthogonal matrix representing rotation. Infinitesimally 
$O_{ij}=\delta_{ij}+\omega_k\varepsilon_{kij}$ which yields

\begin{equation}
 \label{e54}
\delta{\mA}_i(p') \equiv {\mA}'_i(p) - {\mA}_i(p) = -\omega_k\varepsilon_{klj}p^j\partial_l{\mA}_i(p)
+\omega_k\varepsilon_{kij}{\mA}_j(p)
\end{equation}

For rotationally invariant field $ \delta{\mA}_i=0 $ and one finds

\begin{equation}
 \label{e55}
i\varepsilon_{klj}p^j\partial_l{\mA}_i(p) = [{\mA}_i(p), -i\varepsilon_{klj}p^l\partial_j]=i\varepsilon_{kij}{\mA}_j(p)
\end{equation}

In general, $ \delta{\mA}_i(p)\neq0 $. It is obvious that for vector potential generating rotationally invariant magnetic field is invariant up to a gauge transformation. In fact

\begin{equation}
 \label{e56}
\partial_i\delta {\mA}_j(p) - \partial_j\delta {\mA}_i(p) =
-\varepsilon_{ijm} (-\omega_k\varepsilon_{kln}p^n\partial_lB_m + \omega_k\varepsilon_{kml}B_l)
\end{equation}

which vanishes by virtue of eq.(\ref{e54}) if the field configuration $ B_l $ is rotationally invariant. Consequently, the change of vector potential can be compensated by a gauge transformation.
Let us note that eq. (\ref{e55}) may be rewritten in the form 

\begin{equation}
 \label{e57}
 [{\mA}_i(p), \varepsilon_{klj}p^l(-{\mD}_j)]=i\varepsilon_{kij}{\mA}_j(p)
\end{equation}

where $ {\mD}_j $ is given by eq. (\ref{e12}). More generally

\begin{equation}
 \label{e58}
 \delta{\mA}_i(p)= \omega_k \left( i\varepsilon_{kij}{\mA}_j(p) + [{\mA}_i(p), \varepsilon_{klj}p^l(-{\mD}_j)] \right)
\end{equation}

Let us compute

\begin{equation}
 \label{e59}
 i\delta(i{\mD}_i)= \omega_k \left( i\varepsilon_{kij}(i{\mD}_j) + [i{\mD}_i, \varepsilon_{klj}p^l(-i{\mD}_j)] \right)
\end{equation}

One finds

\begin{equation}
 \label{e60}
 i\delta(i{\mD}_i)=i \omega_k \left( p^iB_k-\delta_{ik}p^lB_l \right)
\end{equation}

Using $ B_k=\frac{\lambda p^k}{(p^0)^3} $ eq. (\ref{e60}) becomes

\begin{equation}
 \label{e61}
 i\delta(i{\mD}_i)=i\lambda \omega_k \left( \frac{p^ip^k}{(p^0)^3}-\frac{\delta_{ik}}{p^0} \right)=
 -i\lambda \omega_k \partial_i\left(\frac{p^k}{p^0} \right)
\end{equation}

We see that in order to achieve invariance, $ \delta({\mD}_i)=0 $, one has to compensate rotation by a gauge transformation with gauge parameter $ \lambda \omega_k \frac{p^k}{p^0} $. This amounts to modification of rotation generator by adding $ \lambda \frac{p^k}{p^0} $.

\subsection{Generators in Wu-Yang embedding}

Consider the Skagerstam \cite{b11} description of photon states described in Sec. IV. The generators of the Lorentz group in the picture involving Wu-Yang connection read:

\begin{equation}
 \label{e62}
 \begin{split}
 (J_k)_{mn}=\varepsilon_{kij}(x_i)_{mn}p^j + \frac{p^k}{(p^0)^2}p^l(\mathcal{S}_l)_{mn}\\
 (K_k)_{mn}=\frac{1}{2}\left(p^0(x_k)_{mn}+(x_k)_{mn}p^0\right)
 \end{split}
\end{equation}

where

\begin{equation}
 \label{e63}
 \begin{split}
 &(x_k)_{mn}=\left( i\partial_k + \frac{i}{2} \frac{p^k}{(p^0)^2}\right)\delta_{mn}+
 i\frac{p^m}{(p^0)^2}\delta_{kn} - i\frac{p^n}{(p^0)^2}\delta_{km}\\
 &(\mathcal{S}_l)_{mn}= -i\varepsilon_{lmn} 
 \end{split}
\end{equation}
   
 It is straightforward to check that the action Poincare generators preserves transversality condition $ p^k\varphi_k(p)=0 $.
 Let us first note that the rotation generators can be written in the known form  
   
 \begin{equation}
 \label{e64}
 (J_k)_{mn}=\delta_{mn}\varepsilon_{kjl}p^j(-i\partial_l) + (\mathcal{S}_k)_{mn}
\end{equation} 

On the other hand the boosts read 
   
 \begin{equation}
 \label{e65}
 (K_k)_{mn}=i\delta_{mn}p^0\partial_k+i\frac{p^m}{(p^0)}\delta_{kn}-i\frac{p^n}{(p^0)}\delta_{km}
\end{equation}  
     
\subsection{Nonlinearly transforming coordinates}

The coordinate operators obeying (i)$\div$(iii) can not be constructed unless $ \mid\lambda\mid=1/2 $ 
and both helicities are included. However, for any $ \lambda $ one can find coordinates obeying (i) and (ii); to this end it is sufficient to identify coordinate operator wih covariant derivative (multiplied by imaginary unit). It is also possible to construct the coordinate operator such that (i) and (iii) hold. The price one has to pay is that the transformation rules under rotations become more complicated: the coordinates transform nonlinearly and the action of rotation group linearizes on $ SO(2) $
subgroup of rotations around the direction of standard vector. The relevant construction is quite straightforward \cite{b16} . It is well known that the representations of the Poincare group can be obtained by quantization of the relevant Hamiltonian systems on coadjoint orbit. By Darboux theorem the symplectic form on coadjoint orbit can be put in standard form. The coordinate operators are then the quantum counterpart of Darboux coordinates. There is no point to repeat here the full reasoning \cite{b16}.
However, the final result is quite simple. The resulting coordinate operators differ from the ones obeying (i) and (ii) by a (minus) gauge potential in some fixed gauge. As a result they are basically the covariant derivatives corresponding to pure gauge configurations so they commute. However, the gauge field can at most have an explicit symmetry with respect to the rotations around one axis while the remaining rotations take nonlinear form (which is another way of expressing the property that the effect of rotations can be compensated by a gauge transformation).
   
\par
{\bf Acknowledgments}   
 The author would like to thank Professors Krzysztof Andrzejewski, Joanna Gonera and Cezary Gonera for helpful discussions and useful remarks.\\
This work has been supported in part by the grant 2016/223/B/ST2/00727 of National Science Centre, Poland


\begin{thebibliography}{99}
\bibitem{b1}\textsc{L. Landau, R. Peierls}, Z. Phys. {\bf 69} (1931), 56;\\
 \textsc{V. B. Berestetskii, E. M. Lifshitz, L. P. Pitaevskii}, Relativistic Quantum Theory I, Pergamon Press 1971 

\bibitem{b2}\textsc{T. D. Newton, E. P. Wigner},Rev. Mod. Phys. {\bf 21} (1949), 400;

\bibitem{b3}\textsc{M. H. L. Pryce}, Proc. Roy. Soc. Lond. {\bfseries A195} (1948), 62;

\bibitem{b4}\textsc{A. S. Wightman}, Rev. Mod. Phys. {\bfseries 34} (1962), 845;\\

\bibitem{b5}\textsc{T. F. Jordan, N. Mukunda} Phys. Rev.  {\bfseries 132} (1963), 1842;

\bibitem{b6}\textsc{A. L. Licht}, Journ. Math. Phys. {\bf 7} (1966), 1656;

\bibitem{b7}\textsc{J. M. Jauch, C. Piron}, Helv. Phys. Acta.  {\bfseries 40} (1967), 559;

\bibitem{b8}\textsc{W. O. Amrein}, Helv. Phys. Acta. {\bfseries 42} (1969), 149;

\bibitem{b9}\textsc{M. J. Shirkov}, Theor. Math. Phys.  {\bfseries 42} (1980, 134;

\bibitem{b10}\textsc{S. N. M. Ruijsenaares}, Ann. Phys.  {\bf 137} (1981), 33;

\bibitem{b11}\textsc{B. S. Skagerstam}, Localization of Massless Spinning Particles and the Berry Phase, preprint,  arXiv:hep-th/9210054;

\bibitem{b12}\textsc{E. P. Wigner}, Ann. Math. {\bf 40} (1939), 149;

\bibitem{b22}\textsc{H. Bacry}, Journ. of Phys.  {\bf A14}, (1981), L73

\bibitem{b13}\textsc{B. Ek, B. Nagel}, Journ. Math. Phys. {\bf 25} (1984), 1662;

\bibitem{b23}\textsc{M. Plyushchay}, Nucl. Phys.  {\bf B589}, (2000), 413
 
 \bibitem{b24}\textsc{C. Duval, M. Elbistan, P. Horvathy, P.-M. Zhang}, Phys. Lett.  {\bf B742}, (2015), 322
 
 \bibitem{b25}\textsc{C. Duval}, " A recollection of Souriau's derivation of the Weyl equation via geometric quantization", arXiv:1602.01054   
 
 
\bibitem{b14}\textsc{R. Jackiw}, Invariance, symmetry and periodicity in gauge theories, Proceedings of Schladming Conference: Field and strong interactions, (1980), p. 67
 
\bibitem{b15}\textsc{R. Jackiw, C. Rebbi,}, Phys. Rev. Lett. {\bf 36} (1976), 1116; \\
\textsc{P. Hasenfrantz, G. 'tHooft}, Phys. Rev. Lett. {\bf 36} (1976), 1119;\\
\textsc{A. S. Goldhaber}, Phys. Rev. Lett. {\bf 36} (1976), 1122;

\bibitem{b16}\textsc{K. Andrzejewski, A. Kijanka-Dec, P. Kosi\'nski, P. Ma\'slanka}, Phys. Lett. {\bf B746} (2015), 417;

\bibitem{b17}\textsc{M. A. Stepanov, Y. Yin} , Phys. Rev. Lett. {\bf 109} (2012), 162001;

\bibitem{b18}\textsc{G. 'tHooft}, Nucl. Phys. {\bf B190} (1981), 455;

\bibitem{b19}\textsc{I. Bia\l ynicki-Birula, Z. Bia\l ynicka-Birula}, Quantum Electrodynamics, Pergamon Press,2013 

\bibitem{b20}\textsc{T. T. Wu, C. N. Yang}, Phys. Rev. {\bf D12} (1975), 3843;

\bibitem{b21}\textsc{C. Duval, P. A. Horvathy}, Phys. Rev. {\bf D91} (2015), 045013; 
  
\end{thebibliography}
\end{document}